\documentclass[rmp,twocolumn,showpacs,superscriptaddress]{revtex4-2}

\pdfoutput=1
\usepackage[utf8]{inputenc}
\usepackage[english]{babel}
\usepackage[T1]{fontenc}
\usepackage{amsmath}
\usepackage{hyperref}

\usepackage{tikz}
\usepackage{lipsum}

\pdfoutput=1

\usepackage[normalem]{ulem}
\usepackage{graphicx}
\usepackage{amsmath}
\usepackage{amssymb}
\usepackage{latexsym}
\usepackage{fancyhdr}
\usepackage{array}
\usepackage{amsfonts}
\usepackage{mathrsfs}
\usepackage{mathtools}
\usepackage{color}
\usepackage{verbatim}
\usepackage{physics}
\usepackage[caption=false]{subfig}
\usepackage{natbib}
\usepackage{enumitem}
\usepackage{tcolorbox}
\usepackage{bbold}

\begin{document}

\title{Relational Quantum Mechanics with Cross-Perspective Links Postulate: an Internally Inconsistent Scheme}

\author{Marcin Markiewicz}
\affiliation{International Centre for Theory of Quantum Technologies (ICTQT),
University of Gdansk, 80-308 Gdansk, Poland}

\author{Marek \.Zukowski}
\affiliation{International Centre for Theory of Quantum Technologies (ICTQT),
University of Gdansk, 80-308 Gdansk, Poland}

\date{\today}

\begin{abstract}

We discuss the status of relative facts -- the central concept of Relational Quantum Mechanics (RQM) -- in the context of the new amendment to RQM called cross-perspective links postulate. The new axiom states that by a proper measurement one learns the value of the relative outcome/fact earlier obtained by another observer-system. We discuss a Wigner-Friend-type scenario in which, without cross-perspective links postulate, relative facts have no predictive or causal power, whereas including cross-perspective links makes them effectively hidden variables, which causally  determine outcomes of specific measurements. However, cross-perspective links axiom invalidates the other axiom of RQM, the one which states that in a Wigner-Friend scenario, RQM assigns an entangled state to the Friend and System  after the unitary transformation of their interaction, despite the appearance of the relative fact for the Friend. This quantum mechanical state according to RQM properly describes the situation for Wigner. From this we show  that RQM with cross-perspective links axiom is an internally inconsistent hidden variable theory and therefore cannot be treated as an interpretation of quantum mechanics in any sense.
 
\end{abstract}

\maketitle

\section{Introduction}
In this work we study consequences of  a modification of Relational Quantum Mechanics (RQM) introduced by Adlam and Rovelli \cite{Adlam.22}. The new axiom introduced therein is called \textit{cross-perspective links} and verbatim reads:

\begin{itemize}
\item {\bf “cross-perspective links:”} In a scenario where some 
observer Alice measures a variable $V$ of a system $S$, then 
provided that Alice does not undergo any interactions which 
destroy the information about $V$ stored in Alice’s physical
variables, if Bob subsequently measures the physical variable 
representing Alice’s information about the variable $V$, then 
Bob’s measurement result will match Alice’s measurement result. \footnote{The cross-perspective-links axiom is in the preprint version of \cite{Adlam.22} numbered 4.1, as it is to replace axiom RQM-4. We will be referring to this axiom throughout the paper either as \textit{cross-perspective links} or RQM-4.1}
\end{itemize}
Most importantly it replaces according to the Authors the following axiom:

\begin{itemize}
    \item 
{\bf RQM-4. Relativity of comparisons:} it is meaningless to compare the accounts relative to any
two systems except by invoking a third system relative to which the comparison is made.
\end{itemize}

As RQM is claimed to be an interpretation of quantum mechanics, all rules of quantum mechanics must be applicable within it. It must give predictions which are in tune with the quantum ones. The additional notions introduced in RQM, with respect to quantum formalism, can interpret the nature of  quantum predictions, but cannot change them. Also any theory has to be internally consistent, cannot lead to contradictions. But a theory which is an interpretation of another one must be consistent with the theory it is intended to interpret. Otherwise it is a different theory, with testable predictive discrepancies.

Based on the above we shall assume that quantum mechanical terms, which are used by the followers of RQM, mean what they mean in quantum mechanics, provided that a {\em caveat} is not expressed by the followers of RQM stressing that this or that must be understood in an RQM way.

With this respect we see no difference between quantum understanding of the notion of "state" of a system and the one in RQM. It is a tool for prediction of future results based on earlier facts. \footnote{In \cite{Rovelli.21} one can read: 
"What is then a “quantum state”? In RQM, it is a
bookkeeping of known facts, and a tool for predicting the
probability of unknown facts, on the basis of the available
knowledge. Since it summarises knowledge about
relative facts, the quantum state   of a system (and a
fortiori its density matrix) does not pertain solely to
the system. It pertains also to the other system involved
in the interactions that gave rise to the facts considered
known. Hence it is always a relative state."

 This  {\it might seem} to be an expression of the most "orthodox" approach to the quantum state, as a mathematical tool for encoding predictions for the future possible measurements, based on a preparation procedure, which effectively is also a measurement (or was tested in earlier measurements to give the states in question). Still, as we shall see the phrase "it summarizes knowledge of relative facts" is going against the orthodox approach in which no notion of "relative facts" as defined in RQM exist. Using the RQM parlance \cite{Rovelli.21} "orthodox" quantum mechanics uses only "stable facts".  

Still, another statement in \cite{Rovelli.21} seems to corroborate with quantum mechanics. Namely
"A moment of reflection shows that any quantum state
used in real laboratories, where scientists use quantum
mechanics concretely, is always a relative state. Even a radical believer in a universal quantum state would concede that the $\psi$
  that physicists use in their laboratories to describe a
quantum system is not the hypothetical universal wave
function: it is the relative state, in the sense of Everett,
that describes the properties of the system, relative to
the apparata it is interacting with." However, in "orthodox" quantum mechanics apparata are always assumed to be macroscopic.}
The predictions are probabilistic in quantum mechanics and in RQM. 

The main difference openly spelled out by followers of RQM is that it introduces the notion of "relative facts", which are produced during a (unitary) interaction of two quantum systems. Such relative facts/outcomes do {\em not} appear in the "orthodox" \footnote{There is just one quantum mechanics, based on facts occurring in preparation and measurement apparata. We use the term "orthodox" to inform about this  our standpoint, in the wording of those who are not satisfied with quantum mechanics as it is. Note that "orthodox" quantum mechanics has not been falsified in any experiment, and since 1925 is successfully used for making predictions.} quantum theory. The following quotation might be best in explaining the new notion:
\begin{itemize}
    \item \cite{Rovelli.21} "For relative facts, every interaction can be seen as a
“Copenhagen measurement”, but only for the systems
involved. Any physical system can play the role of the
“Copenhagen observer”, but only for the facts defined
with respect to itself. From this perspective, RQM is
nothing else than a minimal extension of the textbook
Copenhagen interpretation, based on the realisation that
any physical system can play the role of the “observer”
and any interaction can play the role of a “measurement”:
this is not in contradiction with the permanence of interference
through interactions because the “measured”
values are only relative to the interacting systems themselves
and do not affect other physical systems."
\end{itemize}

One should comment here that the word "realisation" should, we think, be replaced here by "RQM assertion". We also cannot agree, that the extension is "minimal". Still, this quotation defines the RQM standpoint.

\subsection{The basis of our analysis}
Our analysis is interpretation neutral. Interpretations usually involve a specific understanding of the notion of the quantum state.
For us a  quantum state is a theory-specific description
of a statistical ensemble of equivalently prepared systems, which allows for statistical (probabilistic) predictions of future measurements, via the Born rule. \footnote{In \cite{peres2002quantum} we read in page 13 "{\em In a strict sense, quantum theory is a set of rules allowing the computation of}
{\bf probabilities} {\em for the outcomes of tests which follow specified preparations.}", and in pages 25-26, one can find: "Note that the word
“state” does not refer to the photon by itself, but to an entire experimental
setup involving macroscopic instruments. [...] The essence of quantum theory is to provide a mathematical representation
of states (that is, of {\em preparation procedures}), together with rules for computing
the probabilities of the various outcomes of any test." } 
The state describes an individual system only as a member of such an ensemble. 
The theory itself is "a set of rules for calculating probabilities for macroscopic detection events,
upon taking into account any previous experimental information", \cite{fuchs2000quantum}. Or if you like, one can use E. P. Wigner's statement (1961) ``the wave function is only a suitable language for describing the body of knowledge - gained by observations - which is relevant for predicting the future behaviour of the system" \cite{Wigner.61}. 
Note that all internally consistent interpretations (not modifications) of quantum mechanics agree with the above. They only add some other properties to the quantum state and/or variables to the theory (hidden causes, hidden ontic states, hidden many worlds, etc.), or to individual members of the ensemble (systems), without any modification of the calculational rules of quantum theory (based on the statistical ensemble approach).

\subsection{The codification of RQM, and its change}

In  \cite{Adlam.22} (pre-cross-perspectives) RQM  assumptions are listed in the form of the following set of axioms. We quote verbatim from \cite{Adlam.22}:
\begin {itemize}
\item
RQM-1. Relative facts: Events, or facts, can happen relative to any physical system.
\item
RQM-2. No hidden variables: unitary quantum mechanics is complete.
\item
RQM-3. Relations are intrinsic: the relation between any two systems $A$ and $B$ is independent
of anything that happens outside these systems’ perspectives.
\item
RQM-4. Relativity of comparisons: it is meaningless to compare the accounts relative to any
two systems except by invoking a third system relative to which the comparison is made.
\item
RQM-5. Measurement: an interaction between two systems results in a correlation within the
interactions between these two systems and a third one; that is, with respect to a third system
$W$, the interaction between the two systems $S$ and $F$ is described by a unitary evolution
that potentially entangles the quantum states of $S$ and $F$.
\item
RQM-6. Internally consistent descriptions: In a scenario where $F$ measures $S$, and $W$ also
measures $S$ in the same basis, and $W$ then interacts with $F$ to ‘check the reading’ of a
pointer variable (i.e., by measuring $F$ in the appropriate ‘pointer basis’), the two values
found are in agreement.
\end{itemize}

As RQM is to be an interpretation of quantum mechanics let us discuss quantum mechanical meaning of these postulates.
\begin{itemize}
\item 
RQM-1. A new notion of event or fact is introduced. In quantum mechanics events, or  facts, happen only in the preparation stage of an experiment (selection of systems so that they are in a specific initial pure or mixed state) and the final measurement (they are the results), and in both cases they are macroscopically accessible. In between we have an evolution, which might be controlled during the experiment or not.
\item 
RQM-2. An unquestionable definition of  hidden variables in quantum mechanics can be put as follows: Hidden variables are any additional parameters or variables which are introduced to the description and are not present in the quantum formalism. Their aim is to causally influence the results of final  measurements. The causal influence does not have to obey relativistic constraints (witness: Bohmian Mechanics).

Concerning unitary quantum mechanics, by this usually people mean quantum mechanics without the collapse postulate. Yes, it is complete provided one uses it correctly. E.g. the unitary description of proper quantum measurements with macroscopically accessible results must  involve decoherence, see e.g. \cite{ZUREK2022}, or in relation to the matters discussed here \cite{Zukowski.21}. The decoherence is due to
the macroscopic nature of measuring devices which produce "outcomes". Decoherence is the way to understand the emergence of classicality of macroscopic objects from the point of view of unitary description of the dynamics of the atomic structures that constitute them \cite{Zurek.82}. This is a modern version of Bohr's insistence on the classical nature of our observation of measurement events in quantum experiments \cite{CAMILLERI201573}.

One can also use a simplified description involving states of memories encoding the results. The states of the memories are not allowed to be in a superposition. In short measurement devices in quantum mechanics and their memories are on the macroscopic side of Heisenberg's Cut.
\item 
RQM-3. This seems to be also applicable to the facts of point 1.
\item 
RQM-4. A purely RQM postulate, having no relation with formalism of quantum mechanics, because it involves relative facts, see RQM-1.
\item 
RQM-5. The first sentence of the axiom is most probably a typo (as after $S$ and $F$ interaction there is no reason for the state of $W$ to be correlated with them). We shall interpret it in the following way (consistent with earlier writings of authors supporting RQM ideas\footnote{See the following statement by Rovelli in \cite{Rovelli.21}: "[...]in the Wigner’s friend scenario, the friend
interacts with a system and a fact is realised with respect
to the friend. But this fact is not realised with respect
to Wigner, who was not involved in the interaction, and
the probability for facts with respect to Wigner (realised
in interactions with Wigner) still includes interference effects.
A fact with respect to the friend does not count as
a fact with respect to Wigner. With respect to Wigner,
it only corresponds to the establishment of an “entanglement”,
namely the expectation of a correlation, between
the friend and the system."}): 

An interaction between two systems $S$ and $F$  is quantum mechanically described by a unitary evolution and in RQM results in an outcome which is a relative fact for these (it is uncertain in RQM whether for both or one of them, but the symbol $F$ suggests the Friend in Wigner's Friend scenario, therefore there seems to be a preference in RQM to think that the fact occurs for $F$ only).  With respect to a third system
$W$ (this seemingly suggests  Wigner), the only consequence of the (unitary) interaction between the two systems $S$ and $F$ 
is that $S$ and $F$ are  entangled after this process.

This can be put in a form of quantum formulas, which give the proper (according to RQM)  state of $S\otimes F$, "with respect to $W$". Namely  after $S$-$F$ interaction producing a fact/outcome for $F$ only, denoted here as $r_F$, related with her RQM measurement of an observable $\hat{O}=\sum_{l=1}^d r_l\ketbra{l}_S$, the state that $W$ faces\footnote{{\em Important}: the state he faces, not the state he thinks that he faces, or ascribes, or must ascribe, or whatever. A quantum state gives proper predictions, no matter whether the observer knows or not the state. Knowledge of a state may influence the action of the observer, but not the measurement results. } is as follows:

\begin{equation}
\label{SAstateForW}
\ket{{for-W}}_{S\otimes F}=\frac{1}{\sqrt{2}}\big( \sum_{l=1}^d c_l\ket{l}_S \ket{l}_F  \big).
\end{equation}
Note that one of $r_l$'s must be $r_F$, but this is not (more, cannot be) encoded in the state $\ket{{for-W}}_{S\otimes F}.$ 

{\em This state is the basis of any quantum prediction for all possible later measurements by W, because such is the role of "states" in quantum theory.} We spell this out, and add an obvious remark: all predictions for W's measurements are based {\em only} on $\ket{{for-W}}_{S\otimes F}$.

\item RQM-6. What is the difference between this postulate and the cross-perspective links one? In RQM-6
$W$ does not have to measure $S$. 


Also, in RQM-6 it is not stated whether the measurement of $S$ by $W$ is simultaneous with the RQM measurement by $F$. $W$ seems to be Wigner in the Wigner-Friend scenario, and this suggests that he measures later. Definitely $W$ cannot act before $F$'s measurement of $S$: if $W$ measures first, then effectively $F $ and $W$ switch their roles.
Thus, if axiom RQM-6 is not to be in  a conflict with axiom RQM-5, and consequently with (\ref{SAstateForW}), one must assume that it applies to simultaneous measurements of $S$ by $F$ and $W$. This is because,  if $W$ acts later than $F$ a conflict with (\ref{SAstateForW}) arises. State $\ket{{for-W}}_{S\otimes F}$ is assumed to properly describe the system $S\otimes F$ before $W$'s action and it does not single out $F$'s relative result/outcome/fact, denoted here by $r_F$ . Therefore, there is no reason whatsoever for $W$ to get result $r_W=r_F$. In other words in RQM a unitary interaction of $F$ with $S$ produces a result $r_F$, but no collapse to 
\begin{equation}
    \ket{r_F}_S \ket{l=r_F}_F  
\end{equation} 
occurs,
as  the superposition  $\ket{{for-W}}_{S\otimes F}$ according to RQM properly describes for $W$ the post-RQM measurement state of $S\otimes F$.
 \footnote{The collapse postulate can be replaced here a by description following quantum theory of measurement based on decoherence \cite{ZUREK2022}. Still, please note that the collapse postulate has not been shown wrong in any experiment, and is an effective way to describe the post-measurement situation (selection of a sub-ensemble of systems which gave a specific measurement result).}

\end{itemize}

\section{No Cross-perspective links: Relative facts vs EPR-type correlations }
In this section  Alice, or $A$, is the Friend. Bob, denoted by $B$ initially plays the role of  Wigner. 
They share a maximally entangled  state of two qubits (say, polarizations of two separate photons):
\begin{equation}\label{EPR}
\ket{ENT_{max}}_S=\sum_{l=0,1}c_l\ket{l}_1\ket{l}_2.\end{equation}
The state  above is in its Schmidt form, thus $c_l$ are real.
In order to remove the ambiguity of the Schmidt basis we assume that  $c_1\neq c_2$, with $c_1c_2\neq 0$, which guarantees that it is unique.

The  labs of Friend and Wigner are separate but do not have to be far away from each other. Two photons reach the labs, one photon per lab (not necessarily at the same time!).
Despite the fact that we have two photons, their polarizations form one system $S=S_1+S_2,$ or more formally $S=S_1\otimes S_2$. Measurement by Alice on her photon $S_1$ constitutes a  measurement of the whole system
$S$ which is degenerate in eigenvalues. Alice makes an RQM measurement on photon 1, and therefore on $S$. The RQM measurement is assumed to be in the Schmidt basis for photon 1, \textit{and we assume that this fact is known to Bob}, in full analogy to the original Wigner-Friend scenario \cite{Wigner.61} in which Friend's measurement basis is known to Wigner.

Thus, according to RQM  Alice obtains a  relative outcome, denoted here as 
${r_a}$. This is one of the possible (eigen-)values of an observable $\hat{A} = \sum_{l=1}^2 r_l \ketbra{l}{l}_1$, where $\ketbra{l}{l}_1$ denotes a projector onto $\ket{l}_1$. Note that the eigenbasis of the observable is the basis for particle $1$, in the Schmidt decomposition of state $\ket{ENT_{max}}_S$. What is the state of $S$ after Alice receives her outcome is undefined in the axioms of RQM.
However RQM is definitive when postulating what is for Wigner the state of $S$ and $A$ after their interaction.
Namely according to RQM (see Eq. \eqref{SAstateForW} and discussion around), Wigner faces an $S+A$ state described by:
\begin{equation}
    \label{SAstateForB}
    \ket{for-B}_{S\otimes A}= \sum_{l=0,1}c_l\ket{l}_1\ket{l}_2 \ket{l}_A.
\end{equation}
Most importantly, according to RQM, this situation is such no matter what was/is the value of Alice's relative outcome $r_a$. Wigner-Bob has no access to $r_a$ without an additional interaction with Alice-Friend or $S$. 

Wigner-Bob receives photon 2, for our argument this is  relativistically objectively after Alice's RQM measurement. He makes a quantum mechanical measurement or an RQM measurement of his photon $S_2$.
Since the quantum state Bob faces, $\ket{for-B}_{S\otimes A}$, contains maximal and complete information for predicting measurements results, according to quantum mechanics, the prediction is  that his  outcomes $r_b$ would appear with probabilities $P(r_b=r_l)=c_l^2$. Most importantly, conditioning on $r_a$ does not change this distribution, since Alice's RQM measurement is not imprinted into the state Bob faces, $\ket{for-B}_{S\otimes A}$. Thus, {\em RQM relative values of Alice and Bob are not EPR correlated.}
This is in stark contrast with a situation, in which Alice performs a proper  quantum mechanical measurement, and not the RQM one. Assume that she gets  a result $r_{a_{QM}}$. Then according to quantum mechanics, if Bob does not have an access to the result, he faces a mixed state:
\begin{equation} \label{post-meas}
    \rho_{S\otimes A}^{post-meas}= \sum_{l=0,1}c_l^2\ket{l}\bra{l}_1\otimes\ket{l}\bra{l}_2\otimes\ket{l}\bra{l}_A.
\end{equation}
However a communication of a message from Alice disclosing $r_{a_{QM}}$ changes Bob's probability assignment from uniform to deterministic. More, as Bob in the Wigner-Friend scenario is aware that Alice is to do her measurements, he additionally knows that even without her message, his result would match hers, $r_b=r_{a_{QM}}.$ This is because the probabilistic mixture reflected in $\rho_{S\otimes A}^{post-meas}$ is only due to him not knowing $r_{a_{QM}}$.

But \textit{if we treat $A\otimes B$ as one system} making an RQM measurement in the product basis $\ket{l}_1\ket{k}_2$ of the two qubits constituting $S$, they get RQM values always correlated $r_a=r_b$, i.e. they are constrained to  $l=k$. The order of performing measurements by $A$ and $B$ is here totally irrelevant.

\textit{Notice  that the freedom of defining of a system (which is perfectly allowed in quantum mechanics),  as the $A\otimes B$ interacting with $S=S_1\otimes S_2$ or $A$ acting on $S_1$ and separately $B$ acting on $S_2$, changes the RQM prediction. In contrast, Quantum mechanics gives an unambiguous  $r_a=r_b$ in both cases.}  

Other remarks: 

\begin{itemize}
    \item If Alice's relative outcome $r_a$ has no conditional consequences for Bob, as above, then it is metaphysical, by which we mean that it does not have any predictive or causal consequences, since conditioning on its value does not affect Bob's probability assignment in a situation in which Alice's and Bob's outcomes are expected to be correlated. In other words existence or non-existence of Alice's outcome, or its actual for Alice value,  has no influence on Bob's observation aimed at repeating her RQM measurement.  
    \footnote{The entangled state  of $S+A$ that Bob predicts/faces after the measurement by $A$ is a basis of prediction for results of his possible actions (not his predictions, simply: predictions).
It is not a state of his mind with no consequences for his possible measurements.
Obviously, if he errs in his description of the state, the outcomes of his actions will generally not be following predictions based on his erroneous belief. In the Wigner-Friend gedanken experiment one should not assume that Bob is stupid.
}
\end{itemize}

\begin{itemize}
    \item If we assume cross-perspective links, then Alice's result is a hidden variable for Bob, which complements the description of his situation. Namely he faces $\ket{for-B}_{S\otimes A}$ {\em and the hidden} unknown to him  $r_a$. It is $r_a$ which \textit{causally} (deterministically !) implies that  his result is to be $r_b=r_a$. Otherwise there is no reason for  $r_b=r_a$ to occur with probability 1, as nothing like this is a consequence of the state faced by Bob, $\ket{for-B}_{S\otimes A}$.
\end{itemize}

\section{RQM with Cross-perspective links: a hidden variable theory}

Here we discuss the consequences of the postulate quoted as cross-perspective links
[item 4.1 of ref. \cite{Adlam.22}], using the wording of  \cite{Adlam.22}.  Below we show  the 
incompatibility of the new postulate (which incorporates relative facts)  with "Wigner-Friend" postulate RQM-5 of RQM and
quantum mechanics.

\textit{Counterargument}:
Let us consider a scenario in which Bob measures "the physical variable representing Alice’s information about the variable $V$" of a system $S$.  In order to indicate the contradiction which arises when one applies the cross-perspective links axiom to this situation let us go step by step through the entire process. The initial state of the system $S$ expressed in eigenbasis of the observable $V$ may be put as:
\begin{equation} \label{SYSTEM}
\ket{system}_{S}=\sum_jc_j\ket{V_j}_S,
\end{equation} 
where kets  $\ket{V_j}_S$ represent eigenstates of the variable $V$. RQM states that from Bob's perspective Alice gets entangled with $S$, and these eigenstates enter the Schmidt decomposition of the resulting entangled state.
Namely Bob faces the following state of the system $S$ and Alice's physical variables:
\begin{equation} \label{BOBS-VIEW}
\ket{for-B}_{S\otimes APV}=\sum_jc_j\ket{V_j}_S\ket{APV_j}_{APV},
\end{equation}
 where $\ket{APV_j}_{APV}$ represent orthogonal states of Alice, associated with different results of the RQM measurement ($APV$ stands in the notation for ``Alice's physical variables"). Here $j=1,2$, but there is no problem to generalize the situation beyond qubits/polarization and have $j=1,2.,...,d$.

Importantly, we shall make a principal assumption: within RQM Bob is not stupid. By this we mean that the state vector  $\ket{for-B}$ is a correct according to RQM Bob's description of the state of the full system $S\otimes APV$, after the interaction between $S$ and $APV$ producing an inaccessible for Bob  RQM relative  outcome. This is a correct description of the situation he faces, not his perhaps wrong assumption. If Bob is replaced by somebody else, say Cecile, then the new agent would face a similar situation, described quantum mechanically in an identical way.\footnote{Note that in quantum mechanics such a unitary interaction involving only $S$ and $APV$ does {\em not} produce any result. In the quantum measurement  theory this is just a pre-measurement, see e.g.  \cite{schlosshauer2007decoherence}.}

Only one of the kets $\ket{APV_j}_{APV}$ is related with Alice's relative fact (outcome) say $V_{j=r_a}$, namely $\ket{APV_{r_a}}_{APV}$,
but this is not a relative fact for Bob, as he faces the  situation described, according to RQM  by \eqref{BOBS-VIEW}. He must take an action to learn more. Assume that he interacts with Alice in such a way that it does not affect the information stored in her physical variables which are related with her RQM measurement of $S$, concerning variable $V$. This means that the interaction must have the property that the evolution leaves states related with possible different results of her measurement intact, $\ket{APV_j}_{APV}\rightarrow \ket{APV_j}_{APV}$. As according to RQM during an RQM  measurement a unitary evolution entangles Bob with the measured object (in this case Alice),  this must lead to the total state of all involved subsystems in the form:
\begin{equation} \label{BOBS-MEAS}
\sum_jc_j\ket{V_j}_S\ket{APV_j}_{APV}\ket{j}_B,
\end{equation}
and then Bob gets a result/outcome,  denoted here as $r_b$,  as a relative RQM fact  or for Wigner-Bob we have a stable fact (RQM with cross-perspective links assumes that these two methods would lead to the same result fixing $j$). The  interaction leading to the state \eqref{BOBS-MEAS} is a Bob-Alice's physical variables entangling process:
\begin{equation} \label{INT}
    \ket{APV_j}_{APV}\ket{initial}_{B}\rightarrow \ket{APV_j}_{APV}\ket{j}_{B},
\end{equation}
which leaves Alice unperturbed, that is $\ket{APV_j}_{APV}\rightarrow \ket{APV_j}_{APV}$. Under such an interaction the reduced density matrix of Alice does not change, it is still the  one implied by $\ket{for-B}_{S\otimes APV}$.  The interaction (\ref{INT}) changes {\em only the state of Bob}.
Applying Born rule to the state \eqref{BOBS-MEAS}
 we get a non-zero probability of Bob having a wrong result:
\begin{equation}
    P(r_b\neq r_a)=\sum_{j\neq r_a}|c_j|^2 \neq 0.
\end{equation}
However, the cross-perspective links axiom assures that in a situation precisely described above "Bob’s measurement result will match Alice’s measurement result", that is we should expect $ P(r_b\neq r_a)=0$. We see that the cross-perspective links axiom leads to a prediction at odds with the quantum mechanical one based on state  $\ket{for-B}_{S\otimes APV}$, which in turn  is defined by axiom RQM-5. An interpretation of quantum mechanics based on RQM with amendment 4.1 is therefore internally contradictory.

The whole discussion
shows that cross-perspective links axiom cannot be merged with quantum mechanical formalism, therefore it can be interpreted only as an introduction of hidden variables to quantum mechanics. Namely in the described scenario, relative fact of Alice is a hidden variable for Bob, as it causally forces outcome of his measurement to be consistent with Alice's. This is not supported by the quantum formalism, if one considers (\ref{INT}) as the proper description of the situation, and contradicts also the axiom RQM-2  on non-existence of hidden variables. Note that relative fact is a specific type of a hidden variable. In contrast to the typical hidden variables discussed in the context of Bell-type scenarios, or in Bohmian mechanics, it is \textit{not} produced during the preparation of initial state of  system(s), but during the RQM measurement. 

All that raises further questions, the most important is the following one. Any physical interaction happens in a finite time interval ("immediate" quantum gates are non-physical idealizations). Therefore one should ask at which stage of the entangling interaction the relative fact is already produced.

\section{\label{Section-proof}Application of cross-perspective links postulate leads to a contradiction in the case of GHZ correlations}

In \cite{Lawrence22}  an inconsistency of the notion of relative facts (outcomes) is proved, provided one assumes the following two principles.  1-LMZ: RQM is an interpretation of quantum mechanics, and therefore especially the Born rule holds in it, and  2-LMZ:
 “if an interpretation of quantum theory introduces
some conceptualization of outcomes of a measurement, then probabilities of these
outcomes must follow the quantum predictions as given by the Born rule.”
In the next subsection we present a modification of the proof from \cite{Lawrence22} which instead of 2-LMZ uses cross-perspective links axiom. Let us however start from a short summary of the original argument presented in  \cite{Lawrence22}.

The gedanken-experiment setup for showing the contradiction \cite{Lawrence22} consists of three qubits {treated as a single compound system $S$} prepared initially in a GHZ state of its constituents, and later measured  in a relativistcally objective  temporally separated sequence by two "observers" $A$ and $B$  in such a way that  the situation as a whole is an operational scheme of  the so-called \textit{extended Wigner-Friend scenario}, introduced in  \cite{Deutsch.85}. Namely, the first ``observer'' $A$ performs  an \textit{RQM measurement} on the 8-dimensional system $S$, in such a way that we have a unitary interaction composed of three commuting unitary  sub-transformations, each acting on a different qubit degree of freedom of the compound system $S$, which is tensor-factorized into such, $S=\otimes_{l=1}^{3}S_l$. Such a compound transformation, as well as its factors,  form  a pre-measurement in quantum mechanics (for a discussion of this concept see e.g. \cite{ZUREK2018}). However, in RQM it is assumed that they  also lead to a ``relative outcome''. In the discussed case the outcome can take one of eight possible non-degenerate values. Further on, observer $B$ (who can be thought of as just another system, or as a \textit{Wigner}, the final observer, from the Wigner-Friend scenario of the extended form by Deutsch) performs a measurement on joint system consisting of qubits and the observer $A$ in a basis, 
which is \textit{effectively complementary} \cite{Zukowski.21} to the one chosen by $A$. Since both observer $A$ performs the RQM measurement on three qubits, and $B$ performs a quantum mechanical measurement on effectively an 8-dimensional system, they obtain outcomes which can be represented as two three-tuples of numbers,  or three-valued sequences $\mathcal A_1, \mathcal A_2, \mathcal A_3$ and $\mathcal B_1, \mathcal B_2, \mathcal B_3$ respectively. We use the Bell-GHZ convention that $\mathcal A_l$ and $\mathcal B_l$ have values $\pm1$. A contradiction is then shown, which can be reduced to a statement that there is no assignment of real-valued outcomes to these   numbers $\mathcal A_1, \mathcal A_2, \mathcal A_3, $ $\mathcal B_1, \mathcal B_2, \mathcal B_3$ if they are to satisfy constraints predicted by quantum mechanics.

 \subsection{Our argument rephrased}

The system $S$ is composed of three qubits, which are all RQM measured by $A$. $S$ is prepared in a GHZ state:
\begin{eqnarray}
&\ket{GHZ}_{S}&\nonumber \\&=
  \frac{1}{\sqrt{2}}\left(\otimes_{m=1}^3  \ket{+1^{(1)}}_{S_m}
    +\otimes_{m=1}^3  \ket{-1^{(1)}}_{S_m} \right).&\nonumber\\
\label{GHZ1}
\end{eqnarray}
A measurement of observables of a tensor factorizable form, like $X=X_1\otimes X_2\otimes X_3$ \textit{can} be treated as a single measurement. This  is consistent with quantum mechanics, see the Appendix. 

We have a sequential measurement. First Alice makes her RQM measurement on each of qubits in the GHZ state, in a basis of the Mermin's argument.
The unitary operation related with this RQM measurement (for each qubit subsystem) is given by
\begin{eqnarray} \label{MEAS-F}
    \hat U_m^{SA}\left(\ket{l^{(3)}}_{S_m}\ket{\textrm{initial}}_{A_m}\right)&=& 
    \ket{l^{(3)}}_{S_m}\ket{l^{(3)}}_{A_m}\nonumber\\
    &\equiv&
    \ket{l^{(3)}}_{SA_m},
    \label{SFbasis3}
\end{eqnarray}
where
\begin{eqnarray} \label{BASIS3}
\ket{\pm1^{(3)}}_{S_m}=\frac{1}{\sqrt{2}}\left(\ket{+1^{(1)}}_{S_m}
\pm i\ket{-1^{(1)}}_{S_m}\right),
\end{eqnarray}
and subscript $A_m$ denotes Alice's physical variables coupled via the interaction to qubit $S_m$.
We use here the usual "Bell" convention for eigenvalues of the associated observable:  their value is $\pm 1$  for the respective eigenstate. Alice gets her relative values $\mathcal A_1, \mathcal A_2, \mathcal A_3$. Each equal to $1$ or $-1$.
Next Bob makes his {\em effectively complementary measurements} of the compound system
$S\otimes A$ in an orthonormal  basis which contains tensor product vectors of the form $\otimes_{m=1}^3\ket{\pm1^{(2)}}_{SA_m}$, where
\begin{eqnarray} \label{BASIS2}
\ket{\pm1^{(2)}}_{SA_m}=\frac{1}{\sqrt{2}}\left(\ket{+1^{(3)}}_{SA_m}
\pm i \ket{-1^{(3)}}_{SA_m}\right).\nonumber\\
\end{eqnarray}
As the state after the first unitary  transformation $\otimes_{m=1}^3 \hat U_m^{SA}$ is expressible in terms of states $\otimes_{m=1}^3\ket{\pm1^{(2)}}_{SA_m},$ one does not have to define the other 56 elements of the basis.
One can simplify the discussion by assuming that $B$ is Wigner, and therefore makes a real quantum mechanical measurement from the very beginning. The outcomes of such a measurement are accessible to any information storing devices in his lab, including brains, as they are stable and amplified to the macroscopic level. 

Following the algebra in \cite{Lawrence22} one can see that
Bob-Wigner's results satisfy the following relation:
\begin{equation}\label{GHZ-CONTR-1}
\mathcal B_1^W \mathcal B_2^W \mathcal B_3^W=1.
\end{equation}
We have added the subscript $W$ to stress that these are "stable outcomes" in RQM parlance.
However the relative results of Alice are potentially readable in the counterfactual situation of Wigner performing on them the procedure postulated in the cross-perspective links axiom RQM-4.1.
If he decides to apply RQM-4.1 to read out $\mathcal A_2$ and $ \mathcal A_3$ they must follow relation with  $\mathcal B_1^W$ of the following form:
 \begin{equation}\label{GHZ-CONTR-2}\mathcal B_1^W \mathcal A_2 \mathcal A_3=-1.
 \end{equation}
 In the alternative counterfactual cases (different pairs of Alice's results checked by procedure RQM-4.1) one obtains the following constraints:
 \begin{eqnarray} 
    \mathcal A_1 \mathcal B_2^W \mathcal A_3=-1,\\
    \mathcal A_1 \mathcal A_2 \mathcal B_3^W=-1. \label{GHZ-CONTR-4}
\end{eqnarray}
Wigner knows that he is taking part in a Friend-type experiment, thus is aware of the "agreed" unitary action of $A$ leading to her relative result which can be expressed in the form of three values $\mathcal A_1, \mathcal A_2, \mathcal A_3$, see the Appendix.  He also knows the initial state of $S$. 

Therefore each of $B$'s qubit-convention values $\mathcal B_i^W$ must take the value of the negative product of $A$'s relative outcomes: $ \mathcal B_i^W=-\mathcal A_j \mathcal A_k$, with $i,j,k$ of different values. Thus, he does not have to perform these counterfactual measurements at all. The products of values of Alice are encoded in his results, as $ \mathcal B_i^W=-\mathcal A_j \mathcal A_k$.
As he sees $\mathcal B^W_1\mathcal B^W_2\mathcal B^W_3=1$, he concludes that $\mathcal A_i$ cannot exist because a trivial algebra gives:
$$\mathcal A_1 \mathcal A_2 \mathcal A_3=\pm i,$$
while their values are supposed to be real, $\mathcal A_i=\pm1$.


\section{Conclusions concerning contradictions between axioms}

One can introduce the following requirement:  a new notion can be  introduced in physics  if  one can show that it has a predictive power (or causal influence on events).

Relative facts without cross-perspective links axiom do not have the required above property of physical notions. They are metaphysical only, whereas  the cross-perspective axiom makes them effectively hidden variables, as they determine, from  Bob-Wigner perspective,  the result of a possible measurement in a specific context. This context is a measurement which is described in cross-perspective links postulate, or in the axiom RQM-6 (if one assumes that $W$ acts objectively after $F$). Note that this does not have to be the context of the actual measurement by Bob. 

\subsection{Contradictions listed}

\begin{itemize}
    \item 
    The relative facts do not exist in quantum description. They are an RQM notion (axiom RQM-1). They have a predictive power by cross-perspective links axiom RQM-4.1, therefore they are hidden variables (axiom RQM-3 cannot hold).
    \item Axioms RQM-5 and cross-perspective links RQM-4.1 are contradictory.
\end{itemize}

Note that the replacement of axiom RQM-4 by RQM-4.1 makes a defence of RQM much more difficult, as such defence is usually based on axiom RQM-4. Axioms RQM-4.1 and RQM-6 corroborate for the case of measurement by $W$ occurring objectively after the RQM measurement by $F$, and it is difficult to understand what for the replacement of RQM-4 by  RQM-4.1 is done.

\section*{Acknowledgements}
MM and MZ are  supported by  Foundation for Polish Science (FNP), IRAP project ICTQT, contract no. 2018/MAB/5, co-financed by EU  Smart Growth Operational Programme. We cordially thank Jay Lawrence for months of intensive discussions.

%
    
\appendix
\section{Observables of compound quantum systems}

In \cite{CBR}, the authors reject our inconsistency proof. The main argument on which this refutation is based on is a statement that the quantum mechanical constraint on the product of outcomes:
\begin{equation}
\label{constrB}
    \mathcal B_1\mathcal{B}_2\mathcal{B}_3=1,
\end{equation}
does not necessarily hold in RQM, since [\cite{CBR}, p. 5]:\\
\textit{One cannot however simply define an observer $B = B_1 \otimes B_2 \otimes B_3$ relative to which constraint \eqref{constrB}
holds, if there is no interaction involving those three systems after their measurements take place.}

Let us comment on this statement. The discussion here is an expansion of the one in \cite{lawrence2023relational}. First of all in quantum mechanics there is no absolute definition of a single or compound system. A tensor product nature of a Hilbert space,  like e.g.  $\mathcal H=\mathcal H'\otimes \mathcal H''$, does not necessarily imply, that (i) the system under consideration is \textit{bipartite}.
Moreover, (ii) sometimes it is not physically possible to separate parts corresponding to two Hilbert spaces, as is the case of so-called  virtual tensor products \cite{Zanardi97, ZanardiVirt01}. In the latter paper one finds a description of a situation, in which a subspace of a multipartite quantum system decomposes into two subspaces, which do not correspond to a physical partition of the system under consideration, nevertheless one can implement measurements on them which mutually commute. To sum up, quantum mechanics allows us to define a \textit{single} measurement of local factorizable observables like $X=X_1\otimes X_2\otimes X_3$, and each $X_l$ acts in a two dimensional Hilbert space $\mathcal H_l^{(2)}$. The spaces constitute the full Hilbert space for the problem via a tensor product $\mathcal H^{(8)}=\otimes_{l=1}^3 \mathcal H^{(2)}_l$. Hilbert spaces $\mathcal H^{(2)}_l$ do not have to describe states of separated qubits. 

Further, we can even change encoding of outcomes, or in other words ascribe other eigenvalues, which overtly are associated with the states of the compound system $S$.  This is a general feature, often omitted in textbook presentations. It is most evident when one considers observables which are unrelated with classical dynamical variables, e.g. related to detection events in arrays of detectors. In our context this can be illustrated in the following way. Let three dichotomic observables $X_i$, where $i=1,2,3,$ have eigenvalues $x_i=\pm 1$, and let each pertain to a different qubit degree of freedom of a compound system of three qubits. That is, we assume that our eight-dimensional Hilbert space is tensor-factorized into three two-dimensional spaces $\mathcal{H}^{(8)}=\mathcal{H}_1^{(2)}\otimes\mathcal{H}_2^{(2)}\otimes\mathcal{H}_3^{(2)}$. Each of the factor observables are of the form 
$X_1=X_1^{(2)}\otimes I_2^{(2)}\otimes I_3^{(2)}$, and similarly are defined  other $X_i$'s. Finally,  observables with subscript $i$ act  only in $\mathcal{H}_i^{(2)}$, and $I_i^{(2)}$ is an identity.
Let us assign new eigenvalues to the outcomes using the following mapping:
$+1 \rightarrow 0, -1 \rightarrow 1 $. The new values are bits.
In this way each three-qubit result $(x_1,x_2,x_3)$ can be transformed to bit values, say $b_n$ such that the following relation holds: $x_n=(-1)^{b_n}$.
The compound result  $(x_1,x_2,x_3)$ gets transformed to: 
$$(x_1,x_2,x_3) \rightarrow (b_1,b_2,b_3).$$
Next we can use a one-to-one map of the bit sequences to integers $v$ from 0 to 7 written down in the binary system, which can be put as  \textbf{$(b_1,b_2,b_3) \rightarrow b_3b_2b_1 (binary).$}
Note that the (decimal) value of the integers reads $v(b_1,b_2,b_3)=\sum_{i=1}^3 2^{i-1}b_i$.

Since RQM aims at being an interpretation, not a modification of quantum mechanics, it must allow for treating measurement of observable $X$ as a single measurement, also in its redefined version with the new eigenvalues $v$. Note that a rejection of such a possibility implies that a definition of an RQM ``observer'' system should demand that {\it it} can be only a prime-dimensional one.


\end{document}